# Leveraging multi-view data without annotations for prostate MRI segmentation: A contrastive approach

Tim Nikolass Lindeijer, Tord Martin Ytredal, Trygve Eftestøl, Tobias Nordström, Fredrik Jäderling, Martin Eklund and Alvaro Fernandez-Quilez

*Abstract*— An accurate prostate delineation and volume characterization can support the clinical assessment of prostate cancer. A large amount of automatic prostate segmentation tools consider exclusively the axial MRI direction in spite of the availability as per acquisition protocols of multi-view data. Further, when multi-view data is exploited, manual annotations and availability at test time for all the views is commonly assumed. In this work, we explore a contrastive approach at training time to leverage multi-view data without annotations and provide flexibility at deployment time in the event of missing views. We propose a triplet encoder and single decoder network based on U-Net, *tU-Net (triplet U-Net)*. Our proposed architecture is able to exploit non-annotated sagittal and coronal views via contrastive learning to improve the segmentation from a volumetric perspective. For that purpose, we introduce the concept of inter-view similarity in the latent space. To guide the training, we combine a dice score loss calculated with respect to the axial view and its manual annotations together with a multi-view contrastive loss. *tU-Net* shows statistical improvement in dice score coefficient (DSC) with respect to only axial view ($91.25 \pm 0.52\%$ compared to $86.40 \pm 1.50\%$, $P<.001$). Sensitivity analysis reveals the volumetric positive impact of the contrastive loss when paired with *tU-Net* ($2.85 \pm 1.34\%$ compared to $3.81 \pm 1.88\%$, $P<.001$). Further, our approach shows good external volumetric generalization in an in-house dataset when tested with multi-view data ($2.76 \pm 1.89\%$ compared to $3.92 \pm 3.31\%$, $P=.002$), showing the feasibility of exploiting non-annotated multi-view data through contrastive learning whilst providing flexibility at deployment in the event of missing views.

## I. INTRODUCTION

Prostate cancer (PCa) is one of the most prevalent cancers among men, with more than one million men diagnosed with PCa worldwide on a yearly basis [1]. Multiparametric magnetic resonance imaging (mp-MRI) is becoming a key component for early PCa detection, staging, treatment planning and surgical intervention [2, 3]. Its recommended use prior to biopsies is expected to lead to a considerable increase in prostate radiological examinations and radiologist workload [4, 5]. Prostate segmentation in MRI is an important task for PCa diagnosis and management, where measures such as the prostate volume are directly dependent on the quality and accuracy of it [6]. Specifically, a wrong characterization of the prostate volume can lead to unnecessary subsequent tissue sampling [7, 8]. Despite its relevance, manual segmentation of prostate MRI suffers from inter and intra-reader variability due to the high heterogeneity of prostate morphology, which coupled with the growing number of examinations has motivated the development of deep learning (DL) automatic segmentation tools [2, 9, 10, 11].

### A. Related work

Deep learning (DL) and, in particular, Convolutional Neural Networks (CNNs) have shown promising performances for prostate MRI whole gland (WG) segmentation [12]. Prostate WG segmentation challenges such as SPIE-AAPM-NCIE Prostate MRI (PROSTATEx) have been an important factor to promote research in the area, by providing public datasets and a controlled space to compare algorithm performances [13]. A wide range of architectures based on CNNs have been proposed for prostate WG segmentation in those challenges, with U-Net and variants such as nn-UNet amongst the most common ones [11, 14, 15, 16].

Whilst there is a considerable amount of automatic available prostate segmentation approaches, a large majority use *only axial T2-weighted (T2w) scans as their only data source at training and evaluation time*. Regions of the prostate such as apex and base are known to be already challenging regions for prostate segmentation tasks and the use of T2w axial data as a unique source approach can lead to even more difficulties in their delineation, due to volumetric artifacts [12, 16]. Further, segmentations from axial volumes suffer from large slice spacing, which can result in step artifacts [16].

Despite the axial high intra-plane resolution, prostate MRI protocols recommend the acquisition of at least another view (typically, sagittal) due to the high anisotropic nature of MRI acquisitions [3, 16]. Even in the case of some acquisition protocols, *three orthogonal views* (axial, sagittal and coronal) are commonly acquired to facilitate prostate MRI interpretation and accurate manual delineation (Figure 1).

Tim Nikolass Lindeijer and Tord Martin Ytredal share first authorship and are with the Department of Electrical Engineering and Computer Science at the University of Stavanger, Stavanger, Norway (tm.ytredal@stud.uis.no and tn.lindeijer@stud.uis.no)

Trygve Eftestøl is also with the Department of Electrical Engineering and Computer Science at the University of Stavanger, Stavanger, Norway (trygve.eftestol@uis.no).

Tobias Nordström is with the department of Medical Epidemiology and Biostatistics, Karolinska Institute, Stockholm, Sweden and also with the department of Molecular Medicine and Surgery, Karolinska Institute, Stockholm, Sweden (tobias.nordstrom@ki.se).

Fredrik Jäderling is with the department of Radiology, Capio S:t Göran Hospital, Stockholm, Sweden and also with the department of Clinical Sciences, Danderyd Hospital, Karolinska Institute, Stockholm, Sweden (fredrik.jaderling@ki.se).

Martin Eklund is with the Department of of Medical Epidemiology and Biostatistics, Karolinska Institute, Stockholm, Sweden (martin.eklund@ki.se)

Alvaro Fernandez-Quilez is the corresponding author and with the department of Electrical Engineering and Computer Science at the University of Stavanger, Stavanger, Norway (alvaro.f.quilez@uis.no) and also with the department of Medical Epidemiology and Biostatistics, Karolinska Institute, Stockholm, Sweden and with Stavanger Medical Imaging Laboratory (SMIL), Department of Radiology, Stavanger University Hospital, Stavanger, Norway.

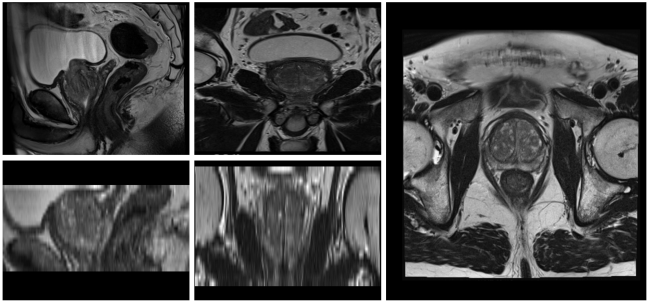

**Fig. 1.** Example of different view scans typically used for segmentation of a patient depicting the volumetric artifacts. On the left side, sagittal and coronal views and on the right side, axial. Bottom row depicts the sagittal and coronal views typically used for segmentation and artifact effect. Upper row depicts the sagittal and coronal scans acquired from the same patient, where the boundaries do not suffer from artifacts.

Different approaches to accommodate and exploit the different views for prostate MRI segmentation have been proposed in the literature. In Cheng et al. [17], the authors proposed an ensemble of three 2D CNNs trained to process *annotated* axial, coronal and sagittal views independently and fusing the outputs of the networks with a *late fusion* approach to extract a high-resolution segmentation. In Meyer et al. [18], the authors proposed a 3D architecture with a multi-stream encoder and *early fusion* mechanism to exploit the 3D nature of prostate MRI and *annotated* axial, coronal and sagittal data. Finally, in Meyer et al. [16], the authors propose a 3D multi-stream architecture able to process anisotropic annotated multi-view MRI with an early fusion mechanism. In all cases, the approaches show a promising performance with improvements in dice score (DSC) when compared to single plane (axial) approaches. Clear improvements are depicted at the base and apex regions in [17, 18] from a relative volume difference (RVD) and DSC perspective.

Although the previous approaches show the benefits of leveraging multi-view MRI prostate data to improve prostate segmentation quality, they *require annotations* for *all* views. That is, the amount of annotations required to train the algorithm is considerably increased with the caveats associated with it in terms of increased time for the radiologist and cost for the health care center. Furthermore, in all cases, *explicit fusion* mechanisms such as *early* or *late* approaches are considered, disregarding similarities in the MRI views from a *conceptual* perspective, where the views can be understood as different representations of the same concept: *the region of interest* (WG). Finally, in all cases, there is an underlying assumption of *availability* of *all the views* used during training at deployment (testing) time or lack of in-depth testing of *realistic* deployment scenarios where the number of available views or annotations can be limited.

### B. Motivation and contributions

In this work, we seek to address the aforementioned limitations and aim to leverage *unannotated* sagittal and coronal T2w prostate MRI together with *annotated* axial view to improve prostate segmentation from a volumetric perspective. For that purpose, we propose a *triplet-like* U-Net architecture and exploit the concept of *interview similarity* in the latent space, avoiding explicit fusion mechanisms. Further, by leveraging the proposed *triplet encoders* we significantly decrease the complexity of the model and allow for flexibility at deployment time, since *the same encoder is used to encode all views*. We evaluate the proposed approach in an *internal cohort* and its generalization ability in an in-house *external cohort*. To the best of our knowledge, *this is the first work that leverages multi-view data without annotation in all but one view* and introduces the concept of *similarity in prostate MRI segmentation*.

The <u>contributions</u> of our work can be summarized as follows:

1. We propose a *triplet-like* U-Net (tU-Net) with shared weights between the encoders able to process multi-view data. Contrary to other approaches, *tU-Net* is designed to provide *flexibility* at deployment time and *account for heterogeneous view-acquisition protocols*.

2. We propose to leverage the concept of *interview similarity* in the latent space at training time, where similarity is understood from the perspective that views are different representations of the same region of interest (ROI): the WG of the prostate.

3. We propose to use *unannotated* sagittal and coronal views, avoiding the costs in terms of time and money associated with obtaining the extra annotations. Contrary to other works, this is the first approach that leverages the multi-view data <u>without annotations</u>.

4. We provide an *internal and external evaluation* of the proposed system. In our internal evaluation, we provide a sensitivity analysis of the different design choices of tU-Net Further, our evaluation focuses on volumetric metrics given their clinical relevance and is presented in terms of different prostate regions. According to and following reproducibility research guidelines, we make our source code available on *GitHub*.

## II. MATERIALS AND METHODS

### A. Data

We use T2-weighted (T2w) axial, sagittal and coronal sequences from the publicly available PROSTATEx dataset [13] as our main data source for model training and development. The nature of the data source is retrospective. The cohort consisted of 204 patients (median age 66 years [range, 48-83 years]) that were under the suspicion of suffering PCa via an initial PSA screening and confirmed through a biopsy. Images were acquired in 2012 with a 3.0-Tesla (T) Siemens scanner with an in-plane resolution of 0.5 mm x 0.5 mm, 3.6 mm slice thickness and with a surface coil. Prostate WG axial segmentations were obtained, at the time of annotation, by two radiology residents and two experienced board-certified radiologists [19].

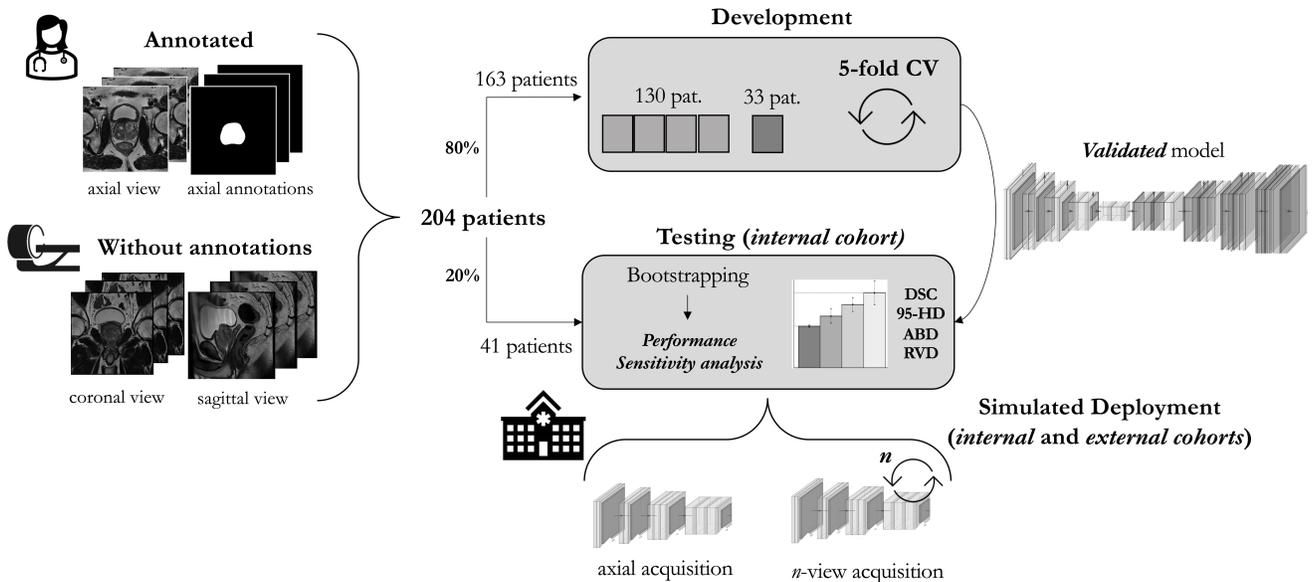

**Fig. 2**. Technical approach to the project: Data splitting, development and design of *tU-Net*, testing and simulation of intended use-cases in the development and external data. DSC stands for dice score coefficient, 95-HD for Hausdorff Distance, ABD for absolute boundary distance and RVD for relative volume difference.

The external evaluation data is from our in-house cohort from Stavanger University Hospital (SUS), Stavanger, Norway. The Regional Committee for Medical and Health Research Ethics (REC Central Norway) approved the use of the dataset (2019/272). The external cohort included a total of 48 patients under the suspicion of PCa based on elevated PSA levels (median age 68 years [range, 49-83]), acquired between 2016 and 2019. The data was acquired with a 3.0 T Philips scanner with an in-plane resolution of 0.5 x 0.5 mm, 3.0 mm slice thickness and with a surface coil. We use T2w axial, sagittal, and coronal sequences together with prostate annotations for the axial view obtained by one radiology resident.

*Pre-processing and data splitting*

We split the main dataset in 80/20% to obtain the train and independent test set, resulting in 163 and 41 patients, respectively. We perform the splitting by patients to ensure no data leakage between the sets. During the splitting, we ensure the different patient views remain matched. Figure 2 depicts the splitting process and technical approach to the project in a schematic way. In the case of the external dataset, we use the whole cohort to test the trained model without re-calibration (out-of-the-box).

To ensure a common space of reference, we apply a linear interpolation resampling to both internal and external evaluation to 0.5 x 0.5 x 3.0 mm, with a slice thickness that is consistent with PI-RADS v2.1 recommendations [3]. Following, we normalize the pixel intensity range with standard pre-processing practices and center-crop the images when necessary [8, 9, 10].

### B. Multi-view contrastive learning for prostate segmentation

To develop *tU-Net*, we adopt a standard U-Net as our base architecture, based on its simplicity and extended use for prostate WG segmentation [11, 12]. In addition to it, we also use nn-UNet and TransUNet for benchmarking purposes, given their wide adoption and state-of-the-art status in the medical image segmentation area [14, 15]. All the architecture variations presented in the work employ a common set-up for training to ensure fair comparisons.

Unless otherwise stated, we use pre-trained models and train for 210 epochs whilst closely monitoring the validation loss and keeping the weights where a minimum value is reached. In the case of the proposed *tU-Net*, we use a simple augmentation policy of image orientation rotation [-60, 60] degrees which is applied in an online fashion at training time with probability p = 0.5 to *all the views*.

In the case of the other architectures, we follow the augmentation policies present in the original works [15]. We choose dice loss as the main contributing loss for training based on its wide acceptance. We use Adam optimizer with a learning rate of rate of 1e-4 and a batch size of 32. Model development and evaluation are carried out on NVIDIA A100 Tensor Core 40GB GPU.

*Triplet encoder U-Net*

Our architecture, *tU-Net (triplet U-Net)*, builds upon a 2D standard U-Net. Figure 3 presents a schematic representation of the architecture. Specifically, we propose the use of *triplet encoders* that share the weights and the architecture design. The lack of connections between the encoder and decoder of the sagittal and coronal views allows us to have a *flexible design* that does *not depend* on the high-resolution information of the rest of branches in the *decoder path*.

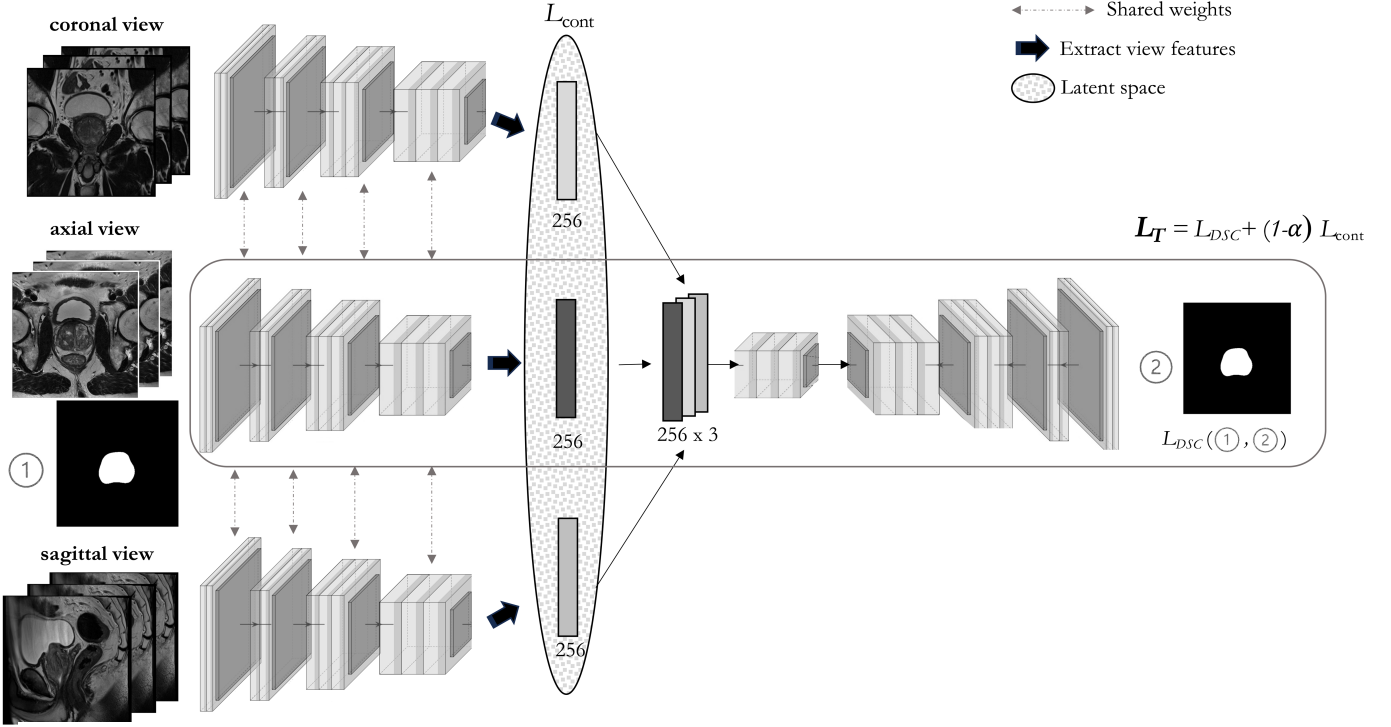

**Fig. 3.** Diagram of the proposed *tU-Net* which accounts for multiple views and incorporates the contrastive loss in the latent space at training time, whilst allowing for a flexible deployment thanks to the bottleneck block.

The design mimics the way a single axial architecture would be trained, aiming for a minimum performance that resembles that of it in the absence of the rest of the views. Whilst adopting that design, we assume a certain trade-off in terms of downgrading of the performance when compared to that of one that connects all views. However, we gain flexibility at deployment time whilst still promoting the usage and information extraction of the views in the back-propagation computation and overall training through the regularization of the latent space.

The *triplet* encoders follow a 5-block structure, where the first blocks are composed of two convolutional layers and a max pooling layer, and the following three of three convolutional layers followed by a max pooling layer. To prevent the network from overfitting the axial data and to force it to learn from the other views, we add dropout layers to the axial view encoder in the last 2 encoder blocks. To ensure the architecture can be used in the presence of only axial views, we map the feature maps of the views to a feature dimension that matches that of the axial view through a *bottleneck layer* (Figure 3, right after the extracted feature maps of the views).

The addition of the bottleneck is a pivotal choice to allow for flexibility at deployment time, where the bottleneck mapping allows re-using the decoder in the presence of a different number of views acquired by the deploying center. The bottleneck block is, again, composed of three convolutional layers together with pooling layers. Following the bottleneck, we have a decoder that mirrors the triplet encoders, resembling a standard U-Net architecture design.

### Regularizing the multi-view latent space: contrastive loss

Inspired by the potential value of the inherent correlation and visual and pixel-level similarities present in prostate multi-view data, we propose the use of a *contrastive loss* ($\mathcal{L}_{cont}$) in the latent space of the *tU-Net* and between the representations obtained from the siamese encoders $f_\theta(\cdot)$. Figure 3 shows a schematic representation of the proposed framework, including the contrastive loss in the latent space. We leverage InfoNCE loss, inspired by other successful approaches [20, 21, 22, 23]:

$$\mathcal{L}^{(v \to u)} = -\log \left( \frac{e^{(\langle v, u \rangle / \tau)}}{\sum_{k=1}^{N} e^{(\langle v_k, u_k \rangle / \tau)}} \right) \quad (1)$$

where $\langle v, u \rangle$ represents the *cosine similarity* between *two encoder* $f_\theta(\cdot)$ representations i.e. $\langle v, u \rangle = v^T u / \|v\| \|u\|$ and $\tau \in \mathbb{R}^+$ is a temperature parameter that can be validated during training.

Intuitively, the contrastive loss aims at predicting $(v, u)$ as a true pair. Minimizing leads to encoders that maximally preserves the mutual information between the pairs. Contrary to other works that use the loss between similar data modalities, our contrastive loss is *asymmetric* for each view. Therefore, given a dataset $D^{AX} = \{(x_i^{AX}, y_i^{AX})\}_{i=1}^{N}$ of $N$ axial T2w sequences and $N$ paired prostate WG segmentation annotations, and the non-annotated datasets $D^{SAG} = \{(x_i^{AX})\}_{i=1}^{N}$ of $N$ sagittal T2w sequences and $D^{COR} = \{(x_i^{COR})\}_{i=1}^{N}$ of $N$ coronal T2w sequences, we adopt the following definition of contrastive loss for our multi-view setting:

$$\mathcal{L}_{cont} = \mathcal{L}^{(x_i^{AX} \to x_i^{SAG})} + \mathcal{L}^{(x_i^{AX} \to x_i^{COR})} + \mathcal{L}^{(x_i^{SAG} \to x_i^{AX})} + \mathcal{L}^{(x_i^{COR} \to x_i^{AX})} \quad (2)$$

where each term is applied to a possible pair of views, *under the assumption that the axial view is the main information carrier* and seeking to define the similarity with respect to it.

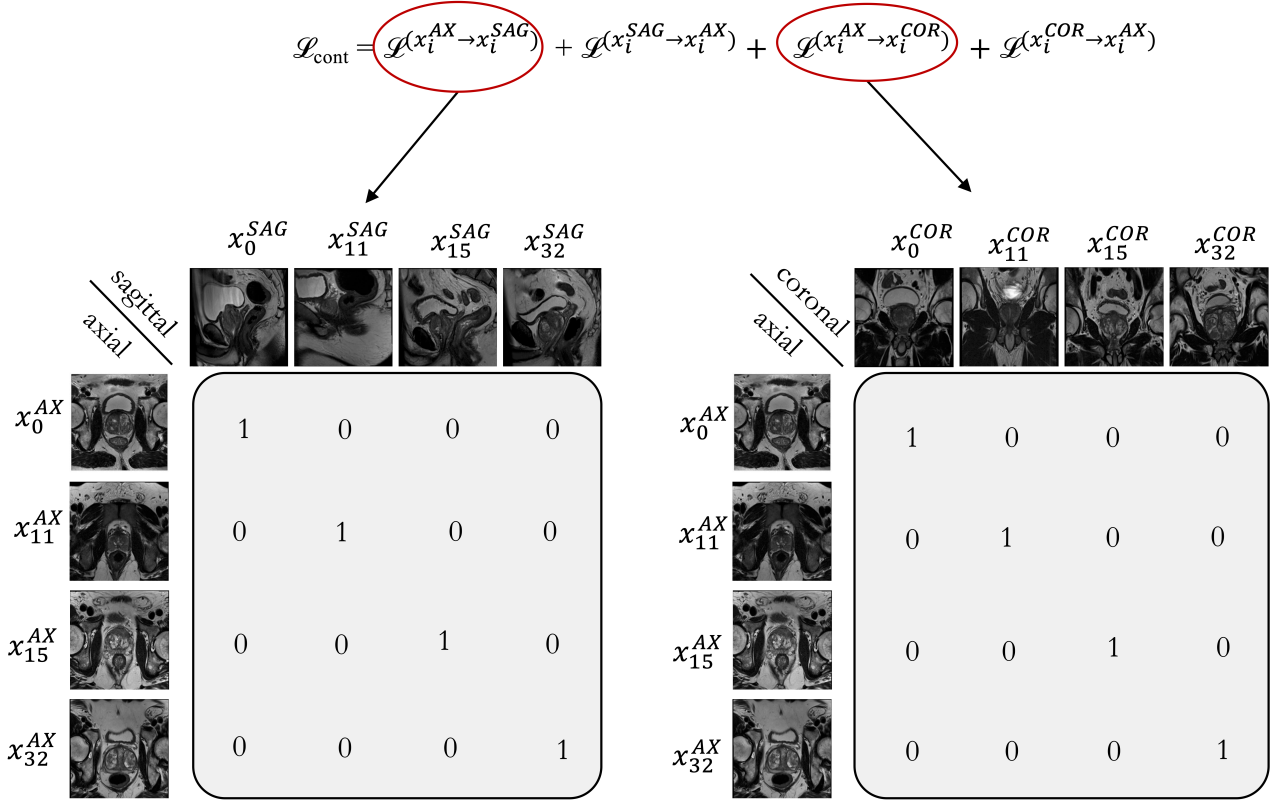

**Fig. 4.** Illustration of the inter-view similarity calculation at training time between slices of the same patient corresponding to different views for $N = 4$ patients. $\mathcal{L}_{cont}$ stands for contrastive loss.

In order to supervise the training of the proposed framework, we weight the previous $\mathcal{L}_{cont}$ and combine it together with a dice loss $\mathcal{L}_{DSC}$, resulting in the final total loss $\mathcal{L}_T$ used to train the system:

$$\mathcal{L}_T = \mathcal{L}_{DSC} + (1 - \alpha)\mathcal{L}_{DSC} \quad (3)$$

where $\alpha$ is an adjustable parameter that controls the strength of the contrastive loss in the training objective.

### Inter-view consistency

In an effort to emulate the way radiologists work, we operate under the assumption that the *axial view* is the *main information* source carrier whilst both sagittal and coronal views can be used to extract complementary and non-redundant information to improve the quality of the segmentation in that view.

In that regard, as shown in Equation 2, our proposed $\mathcal{L}_{cont}$ reflects the assumption by accounting for interactions between the *axial view* and the *sagittal* and *coronal* views. Intuitively speaking, every term in the loss function aims at predicting similar pairs by considering slices of different views *of the same patient* that share *the same spatial position* index to be similar.

We define *similarity* as *spatial similarity* from a ROI perspective i.e. different view slices of the same patient in the same position represent the same ROI. Figure 4 depicts a simple example of the similarity concept described above and how the similarity calculation is performed for a batch of 4 patients.

### Simulating deployment: axial view and multi-view

*tU-Net* accounts for the heterogeneity during acquisition in the number of views of different centers by allowing for flexible deployment scenarios. In particular, we design tU-Net to allow for a deployment scenario in which the *user* can use *tU-Net* out-of-the-box leveraging the 3 orthogonal views or, in the event of having a limited amount of available views as per specifications of the acquisition protocol, to use it out-of-the-box without a significant degradation of the results and to at least, reach the performance of an architecture trained with only axial view.

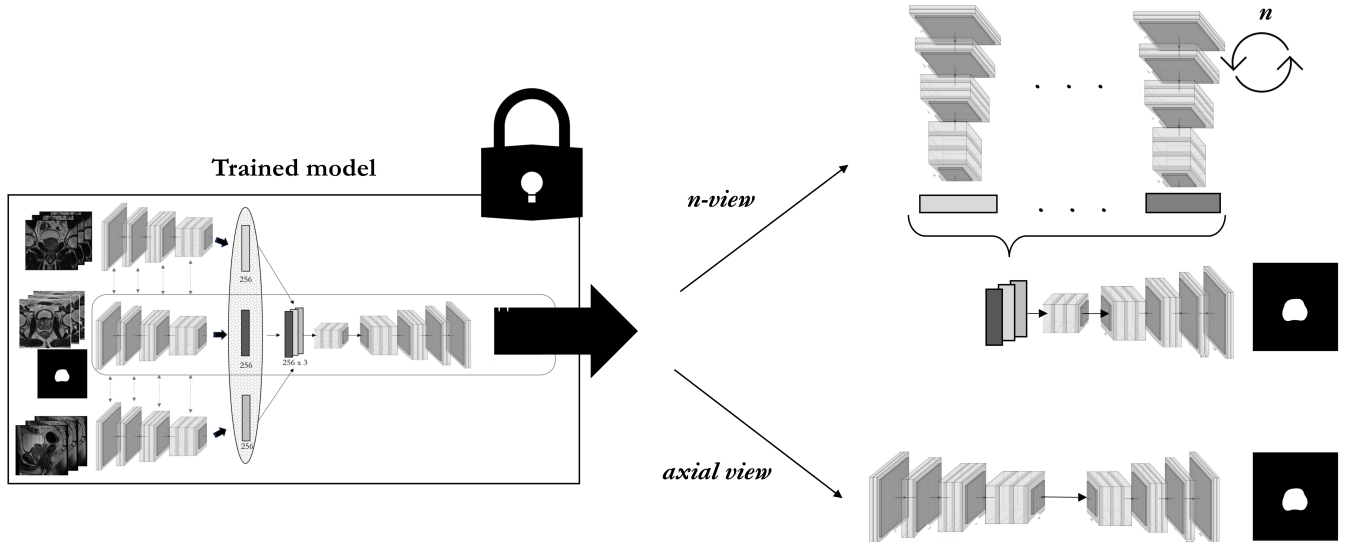

**Fig. 5.** Trained architecture and its flexible use at inference time, where one encoder is re-used for n-views with $n = 1$, 2 or 3.

For that purpose, our architecture relies *exclusively on* the axial view information in the *skip connections* between encoder and decoder, while leveraging the rest of the views for the extraction of non-redundant and additional information, when compared to that of the axial view from a similarity perspective. We accomplish the objective by using triplet encoders with shared weights that combine the multi-view latent space features at training time via a bottleneck that can be adopted or disregarded at deployment time, depending on the number of views available by the deploying center. The choice of design allows to leverage a single encoder at evaluation time to extract the features of all the views with $n$ independent forward pass of the different views of the data whilst reusing the bottleneck architecture (if $n > 1$) and decoder used at training time. Figure 5 depicts the devised deployment scenario and the flexibility of the system.

## III. EVALUATION

We start by presenting the performance results of *tU-Net* trained with three views compared with other single-view state-of–the-art prostate WG segmentation architecture in terms of an *internal testing* with the development set conducted through a 100 bootstrap replicates with repetition. Following, we present a comparison between common multi-view approaches and sU-Net. Additionally, we conduct a *sensitivity analysis* in which we seek to understand the influence of the different design choices: $\alpha$, $\tau$, number of views and $\mathcal{L}_{cont}$.

Finally, we present a *simulated deployment scenario* in which tU-Net is tested in the *development set* and *external set* with a different number of views. In all cases, the results are presented in the form of mean±standard deviation (SD).

Results for the *development phase* and 5-fold cross-validation are presented in Appendix A and results for *tU-Net* trained with different views are presented in Appendix B.

In our attempt to focus on clinically relevant measures, we present results at the *patient level* for different prostate regions of interest: *WG*, *apex*, *mid-gland* and *base*. Each region is assumed to contained in ⅓ of the total number of MRI slices in a given patient sequence [10]. We quantify the quality of the segmentation through Dice Score Coefficient (DSC, %), 95% Hausdorff distance (95-HD, mm), Absolute Boundary Distance (ABD, mm) and Relative Volume Difference (RVD, %). Results are compared quantitatively or by means of t-test or Mann-Whitney U tests for DSC and RVD for WG, where appropriate. A *P* value < .05 is considered to be statistically significant. All statistical tests and models are obtained with Python.

### A. Results

Our 5-fold cross-validation results of *tU-Net* (*3 views* and contrastive loss with $\alpha = 0.25$ and $\tau = 0.07$) showed a better performance for WG than the rest of the benchmarked architectures when internally evaluated during the 5-fold cross-validation: 91.75±0.40 % DSC, 1.06±0.05 mm 95-HD , 0.12 ±0.03mm ABD and 2.80±1.34% RVD. Different *tU-Net* configurations (views, $\alpha$ and $\tau$) were evaluated before reaching the final one and compared based on the 5-fold cross-validation results..

For the remainder of the results, we adopt the previously internally validated through the 5-fold CV *tU-Net* (*3 views* and contrastive loss with $\alpha = 0.25$ and $\tau = 0.07$) and focus *on the test sets* to present the following results. Unless stated otherwise, results are presented for the *internal test cohort of 41 patients from ProstateX*.

| Architecture | Region | Metrics | | | | $p_{DSC}$** | $p_{RVD}$** |
|---|---|---|---|---|---|---|---|
| | | DSC (%) ↑ | 95-HD (mm) ↓ | ABD (mm) ↓ | RVD (%) ↓ | | |
| U-Net | WG | 86.40±1.50 | 1.95±0.31 | 0.30±0.06 | 5.90±3.60 | <.001+ | <.001+ |
| | Apex | 87.50±0.60 | 2.30±0.45 | 0.42±0.11 | 5.30±4.20 | | |
| | Mid | 86.10±2.60 | 1.43±0.37 | 0.20±0.05 | 4.00±3.20 | - | |
| | Base | 87.10±2.00 | 4.39±1.02 | 0.78±0.25 | 6.30±5.30 | | |
| nn-UNet | WG | 89.10±0.42 | 1.41±0.11 | 0.22±0.11 | 4.60±3.10 | <.001+ | <.001+ |
| | Apex | 89.70±0.80 | 2.38±1.12 | 0.41±0.13 | 4.90±2.10 | | |
| | Mid | 88.40±1.20 | 0.96±0.17 | 0.14±0.08 | 3.10±1.40 | - | |
| | Base | 89.82±0.94 | 2.96±1.53 | 0.63±31 | 5.80±3.50 | | |
| TransU-Net | WG | 89.70±1.20 | 1.35±0.18 | 0.20±0.03 | 6.40±3.40 | <.001+ | <.001+ |
| | Apex | 90.10±0.30 | 2.62±1.21 | 0.37±0.12 | 5.60±3.80 | | |
| | Mid | 89.50±1.50 | 0.81±0.36 | 0.13±0.03 | 5.90±1.60 | - | |
| | Base | 90.70±1.00 | 2.91±1.71 | 0.54±0.43 | 6.50±3.40 | | |
| tU-Net | WG | 91.25±0.52 | 1.18±0.11 | 0.16±0.01 | 1.60±1.20 | | |
| | Apex | 91.33±1.47 | 1.44±0.13 | 0.25±0.03 | 3.52±2.22 | | |
| | Mid | 91.11±1.37 | 0.46±0.10 | 0.11±0.03 | 3.19±2.65 | - | |
| | Base | 91.08±1.44 | 2.00±0.43 | 0.30±0.04 | 3.87±2.87 | | |

** Reference is *tU-Net*    +Statistically significant.

**Table 1.** *tU-Net* (trained with 3 views and contrastive loss with $\alpha = 0.25$ and $\tau = 0.07$) segmentation test results compared against single view U-Net, nn-UNet and TransU-Net. Results are presented as mean ± SD.

| Strategy | Region | Metrics | | | | $p_{DSC}$** | $p_{RVD}$** |
|---|---|---|---|---|---|---|---|
| | | DSC (%) ↑ | 95-HD (mm) ↓ | ABD (mm) ↓ | RVD (%) ↓ | | |
| Late concatenation | WG | 87.40±1.51 | 2.05±0.20 | 0.23±0.03 | 5.30±3.20 | <.001+ | <.001+ |
| | Apex | 86.85±1.24 | 2.53±0.24 | 0.51±0.12 | 3.81±2.59 | | |
| | Mid | 86.90±1.25 | 3.53±1.21 | 0.24±0.02 | 3.64±2.58 | - | |
| | Base | 88.45±2.20 | 4.44±0.65 | 0.82±0.34 | 5.58±4.88 | | |
| Early concatenation | WG | 86.70±0.69 | 2.04±0.15 | 0.33±0.05 | 5.42±4.03 | <.001+ | <.001+ |
| | Apex | 86.62±1.31 | 3.35±1.46 | 0.61±0.31 | 5.25±3.51 | | |
| | Mid | 85.18±3.98 | 2.61±0.36 | 0.23±0.03 | 4.84±3.09 | - | |
| | Base | 86.12±2.00 | 4.76±1.32 | 0.88±0.39 | 9.49±6.50 | | |
| tU-Net | WG | 91.25±0.52 | 1.18±0.11 | 0.16±0.01 | 1.60±1.20 | | |
| | Apex | 91.33±1.47 | 1.44±0.13 | 0.25±0.03 | 3.52±2.22 | | |
| | Mid | 91.11±1.37 | 0.46±0.10 | 0.11±0.03 | 3.19±2.65 | - | |
| | Base | 91.08±1.44 | 2.00±0.43 | 0.30±0.04 | 3.87±2.87 | | |

** Reference is *tU-Net*    +Statistically significant.

**Table 2.** *tU-Net* (trained with 3 views and contrastive loss with $\alpha = 0.25$ and $\tau = 0.07$) segmentation test results compared against common prostate WG multi-view existing approaches based on late, early fusion mechanisms and a U-Net architecture. Results are presented as mean ± SD.

### Benchmarking against single view architectures

Table 1 presents the results for the single view axial architectures compared against the proposed *tU-Net*. Significant differences in terms of DSC and RVD are observed between *tU-Net* and the rest of the compared architectures. Furthermore, *tU-Net* shows, qualitatively, improvements for 95-HD and ABD metrics. When assessed by regions, the same trend can be observed, where *tU-Net* significantly outperforms the rest of the architectures, specifically in terms of RVD.

### Performance of tU-Net compared to explicit multi-view fusion mechanisms

Table 2 depicts the results for different common early and late fusion schemes in the same data set-up as *tU-Net*: axial annotations and un-annotated sagittal and coronal views, compared against *tU-Net*. As observed in Table 2, *tU-Net* presents statistically significant differences in terms of DSC and RVD compared to other fusion mechanisms and under the same data conditions.

We can observe from Table 1 that the unannotated nature, in some cases, can result in worse quantitative results for the different segmentation metrics when compared to a standard U-Net. The results reflect the positive impact of the proposed similarity learning scheme, where the network can benefit from the information extracted from the complementary views whilst omitting the need for annotations.

### Sensitivity analysis (alpha, views, and temperature)

The sensitivity analysis is carried out for the different hyperparameters that are pivotal for the design of *tU-Net*: $\alpha$, $\tau$ and the number of views. Specifically, Figure 6 shows the effect of the parameters for a range of predefined values for each parameter at *development time*. That is, the depicted results are obtained during the 5-fold cross-validation training of the model.

As shown in Figure 6, the best configuration based on DSC and RVD consists of *3 views* and contrastive loss with $\alpha = 0.25$ and $\tau = 0.07$, reaching a DSC (%) of 91.75±0.40 and a RVD (%) of 2.80±1.34 at training time. From the results, we can observe the design benefits from a moderate effect of the contrastive loss to regularize the latent space. The interested readers can find the full characterization of the effect of the hyperparameters and a region-based sensitivity analysis in supplementary material.

| Data | Views | Region | Metrics | | | | $p_{DSC}$** | $p_{RVD}$** | |
|---|---|---|---|---|---|---|---|---|---|
| | | | DSC (%) ↑ | 95-HD (mm) ↓ | ABD (mm) ↓ | RVD (%) ↓ | | | |
| ProstateX | axial | WG | 90.07±0.74 | 1.76±0.39 | 0.61±0.10 | 2.01±1.30 | <.001+ | .021+ | internal |
| | axial and coronal | WG | 91.12±0.60 | 1.19±0.13 | 0.16±0.02 | 2.67±1.56 | .70 | <.001+ | |
| | axial and sagittal | WG | 90.79±1.26 | 1.26±0.16 | 0.16±0.02 | 2.58±1.58 | .008+ | <.001+ | |
| | axial, sagittal and coronal | WG | 91.25±0.52 | 1.18±0.11 | 0.16±0.01 | 1.60±1.20 | - | | |
| In-house (SUS) | axial | WG | 66.88±2.48 | 26.14±4.42 | 4.86±0.57 | 3.92±3.31 | <.001+ | .002+ | external |
| | axial and coronal | WG | 75.42±1.67 | 9.16±2.57 | 3.62±0.72 | 3.02±2.85 | <.001+ | .044+ | |
| | axial and sagittal | WG | 75.17±2.50 | 9.77±2.73 | 3.67±0.85 | 2.94±2.78 | <.001+ | .592 | |
| | axial, sagittal and coronal | WG | 79.91±1.18 | 8.01±2.25 | 3.41±0.42 | 2.76±1.89 | - | | |

** Reference is *axial*     +Statistically significant

**Table 3.** *tU-Net* (trained with 3 views and contrastive loss with $\alpha = 0.25$ and $\tau = 0.07$) segmentation test results for WG and different deployment centers (upper row, internal and lower row external). Architecture is applied out-of-the-box. Results are presented as mean ± SD.

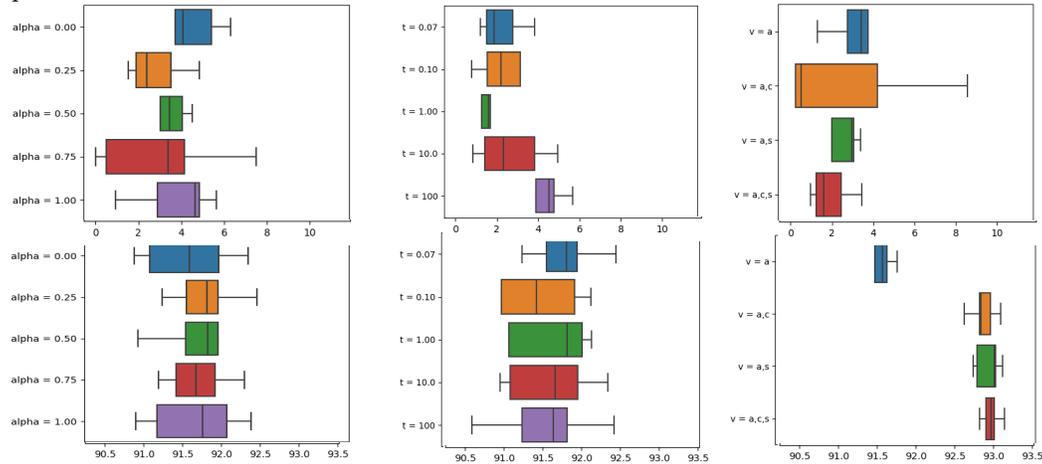

**Fig. 6.** Top row depicts the RVD (%) and bottom row DSC (%) at development time for the different *tU-Net* configurations ($\alpha$, $\tau$ and views).

*Flexible deployment: single view and multi-view scenarios*

Results for the *simulated deployment* scenario are shown in Table 3. Qualitatively, we can observe a decrease in the DSC and RVD performance of *tU-Net* when evaluated in scenarios where missing views are present. In particular, there is a significantly lower performance when the performance is assessed in the event of only axial view, for both the external data evaluation (DSC: 79.91±1.18 *3 views* 66.88±2.48 *axial*, P < .001 RVD: 2.76±1.89 *3 views* 3.92±3.31 *axial* P = .002) and internal data simulated deployment (DSC: 91.25±0.52 *3 views* 90.07 ±0.74 *axial*, P < .001 RVD: 1.60±1.20 *3 views* 2.01±1.30 *axial*, P < .001).

From Table 3, it can be observed that the performance of the system suffers a significantly lower degradation when only either coronal or sagittal views are available, when compared to only axial based on the DSC and RVD values displayed in the different deployment scenarios.

Furthermore, the performance of the system does never go below that of the one observed with only axial view, portraying the usefulness of the additional views in both external evaluation and internal one.

## IV. DISCUSSION

An accurate prostate WG segmentation plays an important role in the diagnosis and management of PCa. Amongst the automatic prostate WG segmentation proposals, a large number focus, exclusively, on the axial view to train the algorithm which can lead to under- performance in challenging regions of the prostate, such as the apex and the base [12, 22].

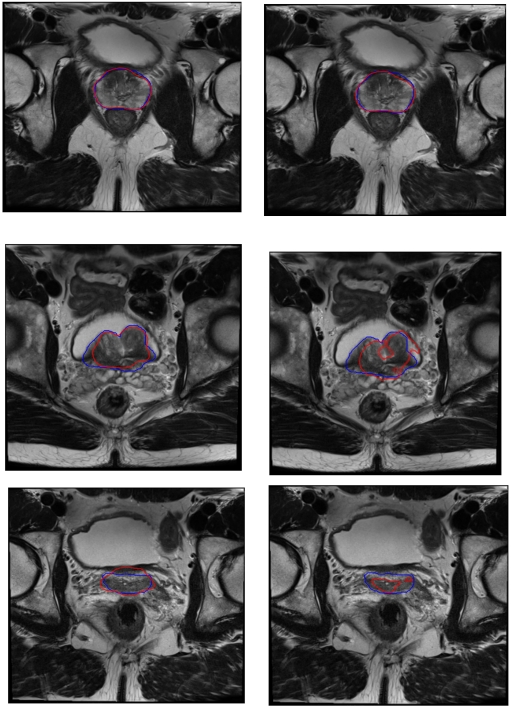

**Fig. 6.** Qualitative segmentation examples of randomly selected patients for axial only (second column) and *tU-Net* with 3 views (first column) for the mid region (mid row) and apex of the prostate (last row). In red, the predicted mask and in blue, the expert annotated mask.

Our results show the significantly positive impact of leveraging the different views when compared to a U-Net trained and tested with only axial view DSC: 91.25±0.52% vs 86.40±1.50%, *P* < .001 and RVD: 1.60±1.20% vs 5.90±3.60%, *P* < .001. Further, we show that *tU-Net* outperforms explicit fusion mechanisms in the presence of non-annotated data DSC: 91.25±0.52% vs 87.40±1.51% (*late fusion*), *P* < .001 and RVD: 1.60±1.20% vs 5.30±4.03% (*late fusion*), *P*< .001.

Additionally, our results show the benefit of including a regularization in the form of similarity loss in the latent space, where our sensitivity analysis shows the benefit of it when compared to not using it in the 5-fold cross-validation results ($\alpha = 1.00$ vs $\alpha = 0.25$): DSC: 91.75±0.44% vs 91.61±0.58%, *P* = .005 and RVD: 2.85±1.34 vs 3.81±1.88, *P*< .001. Particularly, we observe a significant performance improvement in the segmentation apex of the prostate, as depicted qualitatively in Figure 7.

We present a *simulated deployment scenario* where we test the flexibility of the approach in the event of a limited number of views at testing time. The latest is of special relevance, as the acquired number of views in prostate MRI can vary between centers. We show results for two different institutions (internal and external), in which in both, the addition of views benefits the segmentation performance and is never below that of the axial view.

Some solutions have been proposed to leverage the information of sagittal and coronal planes to palliate the aforementioned limitations. Amongst those solutions, Meyer et al. [16] propose an anisotropic multi-stream approach that reaches 92.60±3.00 % DSC and 3.66±2.23 mm 95-HD, surpassing single view axial solutions. Meyer et al. [18] exploit a late fusion mechanism that reaches a 92.10 % DSC and 8.30% RVD.

Whilst prostate WG segmentation multi-planar literature solutions present positive results compared to single view, the authors *require annotations for all the views* at training time. Further, the proposed solutions are *rigid, requiring* the presence of all the views at testing time and not accounting for other possible deployment scenarios where a limited amount of views might be present. On the contrary, our proposed *tU-Net*, a triplet-like U-Net is able to exploit information from all the views without requiring annotations for sagittal and coronal. We leverage a similarity approach and present a flexible framework that allows us to dynamically adapt at inference time depending on the acquisition protocol of the deploying center, as depicted by our results.

## V. CONCLUSION

In conclusion, our research explores the potential similarity learning in enhancing the quality of prostate segmentation from a volumetric perspective, utilizing multi-view data without annotations. We propose a triplet encoder and single decoder segmentation network based on U-Net, capable of leveraging sagittal, coronal, and axial views to improve prostate segmentation without explicit fusion mechanisms. By employing a contrastive loss, we exploit the concept of similarity between different views in the common latent space which guides the training process together with dice score loss based on the axial view annotations.

Our results demonstrate the effectiveness of our proposed methodology in leveraging similarity learning for multi-view, annotation-free prostate segmentation. Given the heterogeneity in the number of acquired views in different centers and rapid adoption of AI tools in radiology, we believe that our approach can facilitate the adoption of the technology by allowing for flexibility at testing time. The results indicate the potential of this approach to improve the accuracy and reliability of prostate segmentation in clinical applications, offering promising avenues for future research and development in the field whilst leveraging non-annotated data.